\documentclass[chi_draft]{sigchi}

\CopyrightYear{2017}
\setcopyright{acmlicensed}
\conferenceinfo{IoT 2017,}{October 22--25, 2017, Linz, Austria}
\isbn{978-1-4503-5318-2/17/10}\acmPrice{\$15.00}
\doi{https://doi.org/10.1145/3131542.3131562}

\usepackage{balance}       % to better equalize the last page
\usepackage{graphics}      % for EPS, load graphicx instead 
\usepackage[T1]{fontenc}   % for umlauts and other diaeresis
\usepackage{txfonts}
\usepackage{mathptmx}
\usepackage[pdflang={en-US},pdftex]{hyperref}
\usepackage{color}
\usepackage{booktabs}
\usepackage{textcomp}

\usepackage{microtype}        % Improved Tracking and Kerning
\usepackage{ccicons}          % Cite your images correctly!
\usepackage{paralist}
\usepackage{enumitem}

\usepackage{tikz}
\usetikzlibrary{shadows,shapes}
\usepackage{pgfplots}
\pgfplotsset{compat=1.10}

\usepackage{xcolor}
\usepackage{colortbl}
\definecolor{Gray}{gray}{0.90}

\usepackage{algorithm}
\usepackage{algorithmicx}
\usepackage{algpseudocode}

\usepackage[textsize=tiny, textwidth=1.3cm, disable]{todonotes}

\newcommand{\Abhishek}[1]{\todo[color=yellow!50, linecolor=black!50]{\textbf{Abhishek}: #1}}
\newcommand{\Aron}[1]{\todo[color=green!40, linecolor=black!50]{\textbf{Aron}: #1}}
\newcommand{\Doug}[1]{\todo[color=red!40, linecolor=black!50]{\textbf{Doug}: #1}}

\newcommand{\circled}[1]{\tikz[baseline=(char.base)]{\node[shape=circle, draw, inner sep=0.5pt, solid] (char) {#1};}}
            
\newcommand{\field}[1]{\texttt{#1}}

\def\plaintitle{Providing Privacy, Safety, and Security in IoT-Based Transactive Energy Systems using Distributed Ledgers}

\def\emptyauthor{}
\def\plainkeywords{Internet of Things; blockchain; transactive energy; privacy; security; transactive microgrid; smart grid; anonymity.}
\makeatletter
\def\url@leostyle{%
  \@ifundefined{selectfont}{
    \def\UrlFont{\sf}
  }{
    \def\UrlFont{\small\bf\ttfamily}
  }}
\makeatother
\def\pprw{8.5in}
\def\pprh{11in}

\definecolor{linkColor}{RGB}{6,125,233}
\hypersetup{%
  pdftitle={\plaintitle},
  pdfauthor={\emptyauthor},
  pdfkeywords={\plainkeywords},
  pdfdisplaydoctitle=true, % For Accessibility
  bookmarksnumbered,
  pdfstartview={FitH},
  colorlinks,
  citecolor=black,
  filecolor=black,
  linkcolor=black,
  urlcolor=linkColor,
  breaklinks=true,
  hypertexnames=false
}

\begin{document}

\allowdisplaybreaks

\title{\plaintitle}

\numberofauthors{4}
\author{%
 \alignauthor{Aron Laszka\\
   \affaddr{Vanderbilt University}\\
%   \affaddr{Nashville, TN}\\
   \email{aron.laszka@vanderbilt.edu}}
 \alignauthor{Abhishek Dubey\\
\affaddr{Vanderbilt University}\\
%   \affaddr{Nashville, TN}\\
   \email{abhishek.dubey@vanderbilt.edu}}
 \alignauthor{Michael Walker\\
\affaddr{Vanderbilt University}\\
%   \affaddr{Nashville, TN}\\
   \email{michael.a.walker.1@vanderbilt.edu}}
    \alignauthor{Doug Schmidt\\
\affaddr{Vanderbilt University}\\
%   \affaddr{Nashville, TN}\\
   \email{d.schmidt@vanderbilt.edu}}
}

\maketitle

\begin{abstract}
Power grids are undergoing major changes due to rapid growth in
renewable energy resources and improvements in battery technology.
While these changes enhance sustainability and efficiency, they also
create significant management challenges as the complexity of power
systems increases.  To tackle these challenges, decentralized
Internet-of-Things (IoT) solutions are emerging, which arrange local
communities into transactive microgrids.  Within a transactive
microgrid, ``prosumers'' (i.e., consumers with energy generation and
storage capabilities) can trade energy with each other, thereby
smoothing the load on the main grid using local supply.  It is hard,
however, to provide security, safety, and privacy in a decentralized
and transactive energy system.  On the one hand, prosumers' personal
information must be protected from their trade partners and the system
operator.  On the other hand, the system must be protected from
careless or malicious trading, which could destabilize the entire
grid.  This paper describes \emph{Privacy-preserving
    Energy Transactions} (PETra), which is a secure and safe solution
for transactive microgrids that enables consumers to trade energy
without sacrificing their privacy.  PETra
builds on distributed ledgers, such as blockchains, and provides
anonymity for communication, bidding, and trading.
\end{abstract}

% TODO!


%!TEX root = paper.tex
\section{Introduction}

% What are transactive energy systems
% What is the key challenge : decentralized information architecture
% What is the related research
% What are the contributions of this paper
% What is the outline of this paper

Power grids are undergoing major changes due to rapid acceleration in
renewable energy resources, such as wind and solar power
\cite{5430489}.
% \cite{EIA2014}
For example, $4,\!143$ megawatts of solar panels were installed in the
third quarter of 2016 \cite{seia}. This capacity is estimated to grow
from 4\% in 2015 to 29\% in 2040 \cite{Randal}. 
%At the same time,
%battery technology costs per kWh have dropped significantly
%\cite{stock2015powerful}, reaching grid parity.
% \cite{bronski2015economics}
The massive integration of renewable energy requires detailed
information and visibility into all aspects of the network, making it
hard to manage, especially in the presence of variable distributed
energy resources \cite{7452738}. A different vision for the future of
power-grid operations is therefore emerging: {\em a decentralized
  system in which local communities are arranged in microgrids}
\cite{rahimi2012transactive}. In this vision, energy generation,
transmission, distribution, and even storage (\emph{e.g.}, electric
vehicles in a community) can be strategically used to balance load and
demand spikes.

Furthering the concept of microgrids, transactive energy models have
been proposed to support the next distribution system evolution
\cite{kok2016society,melton2013gridwise}. Transactive % NOTE: removed cox2013structured to save space
energy is a set of market-based constructs for dynamically balancing
the demand and supply across the electrical infrastructure
\cite{melton2013gridwise}. In this approach, prosumers\footnote{We
  refer to customers as \emph{prosumers} to emphasize that they can
  not only consume energy, but may also produce it.} on the same
feeder (\emph{i.e.}, those sharing a power line link) can operate in
an open market, trading and exchanging generated energy
locally. \emph{Distribution System Operators} (DSOs) can be the
custodians of this market, while still meeting the net demand
\cite{7462854}. For example, the Brooklyn Microgrid
(\url{brooklynmicrogrid.com}) is a peer-to-peer market for locally
generated renewable energy, which was developed by LO3 Energy as a pilot project.

On one hand, transactive energy is a decentralized power system
controls problem \cite{7452738}, requiring strategic microgrid control
to maintain the stability of the community and the utility. On the
other hand, it is a distributed market problem where erroneous---as
well as malicious---transactions can create a gap between demand and
supply, eventually destabilizing the system. In both cases, however,
this system requires a distributed infrastructure comprising smart
meters, feeders, smart inverters, utility substations, the utility
central offices, and the transmission system operator, which must
provide the necessary computation fabric to support the interplay
between the energy control and the fiscal market challenges.
Recently, demand-response systems have been enabled as IoT
applications in smart grids~\cite{Haider2016166}. The transactive grid
described in this paper is the next step in the evolution of energy
systems \cite{collier2017emerging}.

In general, the focus is now on creating a distributed IoT
infrastructure that provides the necessary computation fabric to
support the interplay between energy control and fiscal market
challenges, as shown by Volttron \cite{katipamula2016volttron},
OpenFMB \cite{gunthersmart}, and the Resilient Information
Architecture Platform for Smart Grid (RIAPS)
\cite{eisele2017riaps,Scott2017ICCPS}. For instance, the latter is a
distributed IoT operating system that provides the foundations for all
algorithms, isolates the hardware details from the algorithms, and
provides essential mechanisms for resource management, fault
tolerance, and security. Most of these efforts, however, focus on the
computation and distribution of information, and do not provide the
support required to handle the privacy challenges that arise from the
required information exchange in this decentralized transactive
system.

This paper assumes the existence of a distributed IoT infrastructure
and focuses on the following privacy challenges:
\begin{itemize}[itemsep=0.1\parskip,topsep=-0.75\parskip]
\item \textbf{Leakage of energy usage patterns to other prosumers}
  Since prosumers may purchase energy from each other in a transactive
  microgrid, transactions may inadvertently reveal the prosumers'
  detailed energy usage patterns to other prosumers within the
  microgrid.  Addressing this issue in a decentralized trading system
  is hard as it requires hiding the identities of trade partners from
  each other. In comparison, secure smart metering reveals the
  prosumers' energy usage patterns only to the operator.

\item \textbf{Inference of future states of a prosumer} Transactions
  may reveal the future energy usage of a prosumer, which could be
  used to infer private information.  For example, a smart home may
  know that its inhabitants will go out in the evening (\emph{e.g.},
  by looking at their calendar), and it may trade energy futures
  accordingly in the morning.  Without adequate privacy measures,
  these trades may reveal to other prosumers in the microgrid that the
  inhabitants will not be at home later.  Note that energy futures,
  whose delivery may happen several hours after when the transaction
  is made, can play an important role in predicting and controlling
  microgrid load.  In comparison, smart metering reveals only current
  (or past) usage.

\item \textbf{Personally identifiable information} Transactions and
  energy usage data in a transactive microgrid are much richer sources
  of information than the simple usage data collected by smart meters.
  In particular, the information available in a transactive microgrid
  is a superset of what is available from smart metering and may be
  used to infer personal information, such as risk propensity and
  financial standing.
\end{itemize}
\vspace{0.5\parskip}

Before transactive energy systems can be deployed widely in practice,
the privacy issues described above must be addressed.  Addressing
these issues is hard, however, since solutions must also satisfy
security and safety requirements, which often conflict with privacy
goals.  For example, to prevent a prosumer from destabilizing the
system through careless of malicious energy trading, a transactive
grid must check all of the prosumer's transactions.  In a
decentralized system, these checks require disseminating information,
which could be used to infer the prosumer's future energy consumption.

This paper introduces %an overview of the transactive sequence between
%different components of the systems. We also describe the structure of
%these transactions and show how they can be used to provide
\emph{Privacy-preserving Energy Transactions} (PETra), which is our
distributed-ledger based solution that (1) enables trading energy
futures in a secure and verifiable manner, (2) preserves prosumer
privacy, and (3) enables DSOs to regulate trading and enforce certain
safety rules.
This paper is organized as follows: we first describe
the basic components of a transactive IoT microgrid and formulate
security, safety, and privacy requirements; we next introduce PETra
and describe the transactions and services used to implement it; we
then discuss how it satisfies the security, safety, and privacy
requirements; finally, we 
%compare our work on PETra with 
describe related work
and present concluding remarks.

%!TEX root = paper.tex
%\vspace{-0.15in}
\section{System Model and Requirements}

This section describes a basic system model of transactive IoT
microgrids and formulates security, safety, and privacy requirements.
A microgrid is a collection of prosumers (residential nodes) that are
arranged within the same distribution feeder and support exchange of
power between them. A prosumer node typically includes a smart
inverter and a smart meter, which control the flow of power into and
out of the prosumer. 

A microgrid also typically contains a set of protection nodes that are
responsible for isolating faults on the feeder.  The
\emph{Distribution System Operator} (DSO) operates %a set of
switching nodes to control the connection of the microgrid to the rest
of the distribution system. The DSO is responsible for regulating the
net electric power into and out of the microgrid. Starting from this
model, we next introduce the transactive microgrid model.

\subsection{Transactive Microgrid System Model}
\Abhishek{This is the perfect paragraph to define transactive
  microgrid as a special application of IoT spread over a large
  area. Perhaps we can find a citation that makes this point.}  We
describe a basic system model of decentralized transactive IoT
microgrids.  We discuss the following components: a distributed ledger
for recording transactions, a bid storage service that facilitates
finding trade partners, a microgrid controller for regulating the
microgrid load, and smart meters for measuring the prosumers' energy
production and consumption.

\begin{figure}[h!]
\center
\begin{tikzpicture}[x=8cm, y=0.7cm, font=\small,
  nodeStyle/.style={rounded corners=0.1cm, drop shadow={shadow xshift=0.05cm, shadow yshift=-0.05cm, fill=black}}
]
\draw [nodeStyle, fill=red!10]  (0, 1.65) rectangle    (1, 2.25) node [midway, align=center] {Communication anonymity (e.g., onion routing)};

\draw [nodeStyle, fill=blue!10] (0, 2.4) rectangle    (1, 3.0) node [midway, align=center] {Distributed ledger (e.g., blockchain)};

\draw [nodeStyle, fill=red!10]  (0, 3.15) rectangle (0.45, 4.05) node [midway, align=center] {Transaction anonymity\\[-0.2em](mixing service)};
\draw [nodeStyle, left color=blue!10, right color=red!10] (0.47, 3.15) rectangle (0.72, 4.05) node [midway, align=center] {Bid storage};
\draw [nodeStyle, left color=blue!10, right color=red!10] (0.74, 3.15) rectangle (1, 4.05) node [midway, align=center] {Active\\[-0.2em]smart meter};

\draw [nodeStyle, fill=red!10] (0, 4.2) rectangle (0.49, 5.1) node [midway, align=center] {Anonymous trading\\[-0.2em]workflow for prosumers};
\draw [nodeStyle, fill=blue!10] (0.51, 4.2) rectangle (1, 5.1) node [midway, align=center] {Microgrid controller};
\end{tikzpicture}
\caption{Architecture of a decentralized transactive microgrid with \mbox{PETra}.}
\label{fig:softwareArchitecture}
\end{figure}
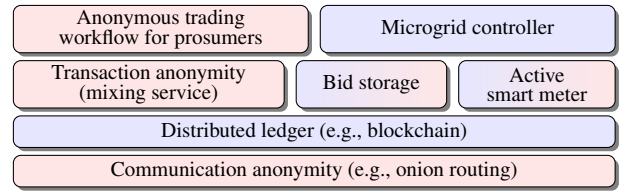

Figure~\ref{fig:softwareArchitecture} shows a decentralized
transactive microgrid with PETra.  
In this figure, components marked in blue are basic elements of the decentralized transactive microgrid,
while components marked in red are added (or extended) by PETra.

\subsubsection{Distributed Ledger}
This ledger permanently stores transactions that specify energy
trades, change regulatory policies for the microgrid, etc.  For
providing security and safety, it is crucial that transactions be
immutable, \emph{i.e.}, after a transaction has been recorded, it
cannot be modified or removed from the ledger.  To enhance fault
tolerance, however, the ledger should also be distributed.

Since a distributed ledger is maintained by multiple nodes, a key
requirement is reaching consensus on which transactions are valid and
stored on the ledger.  Moreover, this consensus must be reached
quickly and reliably, even in the presence of faulty or malicious
(\emph{e.g.}, compromised) ledger nodes.  This paper assumes that a
distributed ledger service is available, but makes no assumptions
about the ledger implementation, such as the particulars of the
consensus algorithm.  In practice, a distributed ledger can be
implemented using, \emph{e.g.}, \emph{blockchains} with proof-of-stake
consensus or a practical Byzantine fault tolerance
algorithm~\cite{castro1999practical}.

\subsubsection{Bid Storage Service}
Although prosumers trade energy with each other directly (\emph{i.e.},
without a middleman), for the sake of scalability, we need a service
that enables prosumers to find trade partners.  We assume that there
is a bid storage service that allows prosumers to post and read energy
\emph{bids} and \emph{asks}.\footnote{A \emph{bid} is an offer to buy
  at a certain price, while an \emph{ask} is an offer to sell at a
  certain price.}  This service relieves prosumers from contacting a
large number of potential trade partners since they only communicate
with the service to discover trade partners.  To enhance scalability
and reliability, this service can also be implemented in a distributed
manner, using multiple nodes.

\subsubsection{Microgrid Controller (Distribution System Operator)}
We assume the existence of a controller at the DSO level that
regulates the total load that the microgrid should present to the
distribution system.  The controller first predicts load in the
microgrid based on (1) bids and asks in the bid storage and (2)
outstanding energy trades in the ledger.  By combining this
information with the prediction for the rest of the grid, the
controller produces a control signal that specifies how much the
microgrid load should be decreased or increased.  Based on this
signal, the controller then updates the price policy for the microgrid
to influence energy production and consumption.  We also assume the
presence of a secondary controller that balances voltage and frequency
in the microgrid.

\subsubsection{Smart Meters}
To measure the prosumers' energy production and consumption, a smart
meter must be deployed at each prosumer.  In practice, these smart
meters must be tamper resistant to prevent prosumers from ``stealing
electricity'' by tampering with their meters.  After a smart meter has
measured the net amount of energy consumed by the prosumer in some
time interval, it can send this information to the DSO for billing
purposes.

\subsection{Requirements}
We now discuss the security, safety, and privacy requirements that
must be satisfied by a transactive energy IoT system.

\subsubsection{Security}
Security requirements ensure primarily that prosumers are billed
correctly, but they also provide necessary prerequisite properties for
safety.
More specifically, they require that
\begin{itemize}[noitemsep,topsep=-\parskip]
\item prosumers are billed correctly based on the energy prices set by
  the DSO, their energy trades, and their actual energy production and
  consumption measured by the smart meters,
\item prosumers or outside attackers cannot change microgrid
  regulatory policies that are set by the DSO, 
\item prosumers cannot back out of trades unilaterally, and they
  cannot tamper with other prosumers' trading or bidding,
\item financial and physical impact of compromised or faulty nodes is
  limited, and nodes can be banned by the DSO. 
\end{itemize}

\subsubsection{Safety}
A careless or malicious prosumer may destabilize the grid by promising
to produce (or consume) a large amount of energy, but failing to actually
produce (or consume) it.  A significant difference between promised and
actual energy production (or consumption) can result in a large gap
between the aggregate production and consumption of the microgrid.
A large gap threatens the stability of not only the microgrid but also the main
power grid.  Therefore, prosumers should not be able to trade large
amounts of energy that they are unlikely to deliver.
Specifically, we require that 
\begin{itemize}[noitemsep,topsep=-\parskip]
\item the net amount of energy sold (or bought) by a prosumer is upper
  bounded (by a limit set by the DSO), where the net amount of
  energy sold is the difference between the amount of energy sold and
  bought by the prosumer, and the net amount of energy bought is
  defined analogously. %,
\Aron{If we have space, add back the safety requirement on the bid storage (and also add back the relevant parts of the description of the trading workflow and the ``analysis.''}
%\item the energy bids and asks posted by a prosumer are limited in a similar way.
\end{itemize}
In practice, the DSO can set the limits based on the prosumers' production and consumption capacities.

\subsubsection{Privacy} 
Privacy requirements ensure that the prosumers' privacy is not
compromised when they participate in energy trading.  We use
non-transactive smart metering as a baseline, and we require that
the transactive system does not leak any additional information
compared to this baseline.  More specifically, we require that
\begin{itemize}[noitemsep,topsep=-\parskip]
\item only the corresponding smart meter and the DSO may gain
  information regarding the amount of energy produced, consumed,
  bought, or sold by a prosumer,\footnote{Although this requirement is
    impossible to satisfy if all other prosumers may collude against
    one target, we can assume that the majority of prosumers are
    non-colluding.}
\item only the prosumer may know which bids and asks it has posted,
  and no one can know who traded energy with whom.
\end{itemize}

%!TEX root = paper.tex
\section{Privacy-preserving Energy Transactions}

This section describes \emph{Privacy-preserving Energy Transactions}
(PETra), which is our solution for providing privacy to prosumers in a
transactive energy IoT system, without compromising grid safety and
security.

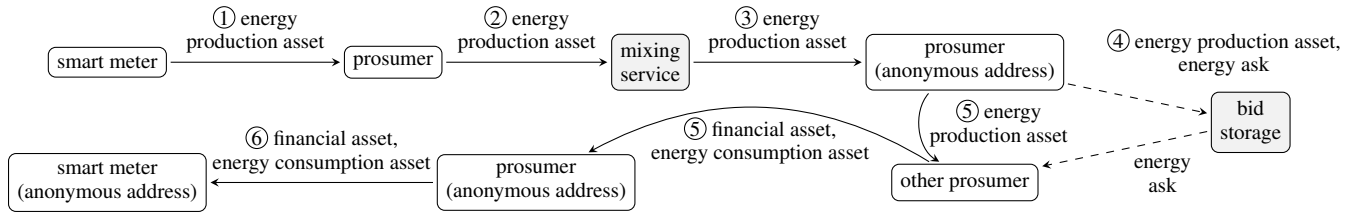
\begin{figure*}[h]
\centering
\begin{tikzpicture}[x=3.8cm, y=1.6cm, font=\small,
  system/.style={draw, align=center, rounded corners=0.1cm, fill=black!5},
  entity/.style={draw, align=center, rounded corners=0.1cm},
  asset/.style={midway, align=center},
  transfer/.style={->, >=stealth, shorten <=0.05cm, shorten >=0.05cm},
]
\node[entity] (smartmeter) at (0, 1) {smart meter};
\node[entity] (prosumer1) at (1, 1) {prosumer};
\node[system] (mixing1) at (1.9, 1) {mixing\\service};
\node[entity] (prosumer2) at (3, 1) {prosumer\\(anonymous address)};
\node[system] (bidstorage) at (4, 0.5) {bid\\storage};
\node[entity] (partner) at (3, 0) {other prosumer};
\node[entity] (prosumer3) at (1.5, 0) {prosumer\\(anonymous address)};
\node[entity] (smartmeter2) at (0, 0) {smart meter\\(anonymous address)};

\draw[transfer] (smartmeter) -- node [asset, above] {\circled{1} energy\\production asset} (prosumer1);
\draw[transfer] (prosumer1) -- node [asset, above] {\circled{2} energy\\production asset} (mixing1);
\draw[transfer] (mixing1) -- node [asset, above] {\circled{3} energy\\production asset} (prosumer2);
\draw[transfer, dashed] (prosumer2) -- node [asset, above right, xshift=-1.5em, yshift=0.5em] {\circled{4} energy production asset,\\energy ask} (bidstorage);
\draw[transfer, dashed] (bidstorage) -- node [asset, below right] {energy\\ask} (partner);
\draw[transfer, bend right=30] (partner) to node [asset, below] {\circled{5} financial asset,\\energy consumption asset} (prosumer3);
\draw[transfer, bend right=50] (prosumer2) to node [asset, right, yshift=0.25em] {\circled{5} energy\\production asset} (partner);
\draw[transfer] (prosumer3) -- node [asset, above] {\circled{6} financial asset,\\
energy consumption asset} (smartmeter2);
\end{tikzpicture}
\caption{Simplified overview of the flow of assets from the perspective of a prosumer who sells energy.
Note that to prevent de-anonymization, a prosumer should use multiple addresses and multiple rounds of mixing, which we have omitted from the figure for clarity of presentation.}
\label{fig:sellFlow}
%\vspace{-0.1in}
\end{figure*}

\subsection{Overview of the Trading Workflow}
We now provide a semi-formal description of the energy trading
workflow from the prosumers' perspective.  Subsequent subsections
describe the assets, transactions, and services used for trading in
more detail.

\subsubsection{Energy Selling Workflow}
Consider a prosumer who wishes to sell energy to another prosumer, as
shown in Figure~\ref{fig:sellFlow}.  \Abhishek{We need to mention what is the
  expected deadline by which this sequence of task should finish.}  As
its first step, the prosumer withdraws an \emph{energy production
  asset} from its smart meter.  An energy production asset represents
a permission to sell a certain amount of energy, and it is used to
enforce safety requirements.  If the prosumer has sufficient unsold
production capacity, the smart meter creates and transfers a
production asset to the prosumer using a \emph{smart meter
  transaction} \circled{1}, which is recorded on the distributed
ledger.

At this point, the production asset can still be traced back to the
prosumer since the ledger is public.  To achieve anonymity, the
prosumer uses a \emph{mixing service}, which could be implemented as a
decentralized protocol, such as
CoinShuffle~\cite{ruffing2014coinshuffle} or
Xim~\cite{bissias2014sybil}.  The prosumer transfers the production
asset to the mixing service using an \emph{energy and financial
  transaction} \circled{2}, which is also recorded on the distributed
ledger.  In turn, the mixing service transfers the production asset to
an \emph{anonymous address} \circled{3}, which is randomly generated
and controlled by the prosumer.\footnote{The concept of \emph{address}
  varies between distributed ledgers, but PETra could be implemented
  using any popular blockchain, such as Bitcoin and Ethereum.
  Specifically, we use the term address to denote a possible
  destination for asset transfers.  Assets that have been transferred
  to an address can be used only by someone who ``controls'' the
  address (typically, the one who generated it), which usually means
  knowing a private key that corresponds to the address.}  Since the
mixing service transfers assets from multiple prosumers to multiple
anonymous addresses at the same time, and the anonymous addresses were
generated at random by the prosumers, the assets cannot be traced back
to the original prosumers after mixing.\footnote{Note that prosumers
  should divide their assets between multiple anonymous addresses;
  otherwise, each asset might be traced back to its prosumer based on
  the amount of energy that it contains.}

The prosumer can now engage in energy trading anonymously.  To find a
trade partner, it can either post an \emph{energy ask} on the bid
storage, or simply search the storage for an acceptable \emph{energy
  bid}.  To post an energy ask, the prosumer first proves to the
storage service---without revealing its original identity---that it
owns a production asset stored at an anonymous address.  
\Aron{This sentence was added because we removed the relevant safety requirement.}
Proving ownership prevents the prosumer from ``spamming'' the storage service
with bogus asks.  The prosumer can then post the energy ask
\circled{4}, which contains an anonymous communication
identifier\footnote{We discuss communication anonymity later.}, a
price, and a reference to the production asset.  If another prosumer,
who is seeking to buy energy, finds the ask acceptable it can contact
the seller using the communication identifier included in the ask.

The seller and buyer can execute the trade by creating an energy and
financial transaction together \circled{5}, and recording it on the
ledger.  This transaction transfers the production asset from the
seller to the buyer, and a \emph{financial asset} and an \emph{energy
  consumption asset} from the buyer to the seller.  A financial asset
represents a certain amount of money, while a consumption asset
represents a permission to buy a certain amount of energy, which is
used to enforce safety requirements similarly to production assets.

Finally, the selling prosumer deposits the financial and consumption
assets to its smart meter using an energy and financial transaction.
To ensure that the prosumer remains anonymous, it transfers the assets
to an anonymous address that is randomly generated and controlled by
the smart meter \circled{6}.  Once the smart meter has received the
assets, it credits the financial amount to and deducts the energy
amount from the prosumer for billing purposes.  To enforce safety
requirements, the prosumer is required to always deposit the same
amount of consumption assets as the amount of production assets
withdrawn at the beginning; otherwise, unaccounted assets could be
used to trade excessive amounts.

\subsubsection{Energy Buying Workflow}
Consider a prosumer who would like to buy energy from another
prosumer.  Since the trading workflow is very similar to the case of
the selling prosumer, we will discuss only the differences.  In the
first step, the prosumer tries to withdraw a financial asset and an
energy consumption asset from its smart meter.  If the prosumer has
the consumption capacity and good financial standing, the smart meter
transfers the assets to the prosumer and adds the financial amount to
the prosumer's bill.

After transferring the assets through a mixing service, the prosumer
is ready to post an energy bid on the bid service.  To do so, it first
proves the ownership of both the financial asset and the consumption
asset to the service, and then posts the energy bid, which includes an
anonymous communication identifier.  If a partner is found, the trade
is executed as described above, with the prosumer playing the role of
the buyer this time.

Finally, the prosumer deposits the purchased energy production asset
to the anonymous address of its smart meter, which credits the energy
amount to the prosumer, for billing purposes.  Note that if the
prosumer has not spent all of its financial assets, then the remainder
may also be deposited back to the smart meter.

%!TEX root = paper.tex
\subsection{Transactions}

The previous subsection gave an overview of how PETra uses
transactions in the trading workflow to transfer various assets.  We
now describe the format of these transactions, as well as the rules
that they have to satisfy to be valid and recorded on the ledger.  We
also introduce and detail regulatory transactions, which the DSO uses
to regulate the microgrid.

\subsubsection{Timing}
The ability to specify points or intervals in time is crucial.  For
example, control signals specify how the microgrid load should change
at certain points in time, energy trades specify when energy will be
consumed or produced, etc.  To facilitate representing signals and
transactions, we divide time into fixed-length intervals, and specify
points or periods in time using these discrete timesteps.  The length
of the time interval is determined based on the timing assumptions of
the physical power system.  For example, the default length of the
time interval may be 4 seconds, which corresponds to how frequently
the control signal of the DSO typically changes.  \Abhishek{We need to
  add citation here. I will add that tomorrow.}  \Abhishek{What about
  the deadline within which the transactions should finish? Do we need
  to say anything here?}  \Aron{Ideally, we should discuss the timing
  constraints of the ledger (probably when we introduce it), but we
  would first need to make space for this discussion.}

\subsubsection{Assets}
Before we can discuss transactions, we need to define the format of
the three types of assets that these transactions may transfer.
First, an \emph{energy production asset} (EPA) is defined by
\begin{itemize}[noitemsep,topsep=-\parskip]
\item \field{power}: non-negative amount of power to be produced (for example, measured in watts),
\item \field{start}: first time interval in which energy is to be produced,
\item \field{end}: last time interval in which energy is to be produced.
\end{itemize}
\vspace{\parskip} Second, an \emph{energy consumption asset} (ECA) is
defined by the same fields.  For this asset, however, the fields
define energy consumption instead of production.  Finally, a
\emph{financial asset} (FA) is defined by a single non-negative number
\field{amount}, which can be denominated in either a fiat currency
(\emph{e.g.}, Euros or US dollars) or a cryptocurrency (\emph{e.g.},
Bitcoin or Ether).

\subsubsection{Energy and Financial Transactions}
Energy and financial transactions transfer energy and financial assets
from one address to another.  Prosumers can use these transactions for
multiple purposes, {\em e.g.}, to trade energy by exchanging assets
with other prosumers, to prove to the bid storage service that they
possess an asset, to hide their identity by transferring assets to and
from mixing services, and to deposit assets at their smart meter.
An energy and financial transaction contains the following fields:
\begin{itemize}[noitemsep,topsep=-\parskip]
\item \field{EPA\_in}: list of EPA inputs, each of which is defined by
\begin{itemize}[leftmargin=0.5em,nosep]
\item \field{out}: reference to an EPA output of a previous transaction,
\item \field{sig}: signature of the referenced output's address,
\end{itemize}
\item \field{ECA\_in}: list of ECA inputs (i.e., list of (\field{out}, \field{sig}) pairs), %, each of which is defined by
%\begin{itemize}[leftmargin=0.5em,nosep]
%\item \field{out}: reference to an ECA output of a previous transaction,
%\item \field{sig}: signature of the referenced output's address,
%\end{itemize}
\item \field{FA\_in}: list of FA inputs (i.e., list of (\field{out}, \field{sig}) pairs), %, each of which is defined by
%\begin{itemize}[leftmargin=0.5em,nosep]
%\item \field{out}: reference to an FA output of a previous transaction,
%\item \field{sig}: signature of the referenced output's address,
%\end{itemize}
\item \field{EPA\_out}: list of EPA outputs, each of which is defined by
\begin{itemize}[leftmargin=0.5em,nosep]
\item \field{EPA}: an energy production asset,
\item \field{address}: address to which EPA is transferred,
\end{itemize}
\item \field{ECA\_out}: list of ECA outputs (i.e., (\field{ECA}, \field{address}) pairs), %, each of which is defined by
%\begin{itemize}[leftmargin=0.5em,nosep]
%\item \field{ECA}: an energy consumption asset,
%\item \field{address}: address to which ECA is transferred,
%\end{itemize}
\item \field{FA\_out}: list of FA outputs (i.e., (\field{ECA}, \field{address}) pairs). %, each of which is defined by
%\begin{itemize}[leftmargin=0.5em,nosep]
%\item \field{FA}: a financial asset,
%\item \field{address}: address to which FA is transferred.
%\end{itemize}
\end{itemize}
\vspace{0.5\parskip} This transaction transfers the assets specified
in the input lists to the addresses specified in the output lists.
Input and output lists may differ in length, so one asset may be
divided into multiple assets, and multiple assets may be combined into
one.

An energy and financial transaction is valid (and can be recorded on
the ledger) if the following three conditions hold.
\begin{itemize}[noitemsep,topsep=-\parskip]
\item None of the outputs referenced by the inputs have been spent by
  a transaction that has been recorded on the ledger.
\item All signatures are valid, which ensures that an asset can be
  transferred only by its current owner. 
\item For each asset type (and for each timestep), the sums of the
input and output assets are equal.  For example, in the case of
energy production assets, the condition is
\begin{align*}
 \forall t: ~ & \textstyle \sum_{\substack{out ~ \in ~ \field{EPA\_out}:\\out.\field{EPA}.\field{start} \, \leq \, t \, \leq \, out.\field{EPA}.\field{end}}} out.\field{EPA}.\field{power} \nonumber \\
 & \textstyle = \sum_{\substack{in ~ \in ~ \field{EPA\_in}:\\in.\field{out}.\field{EPA}.\field{start} \, \leq \, t \, \leq \, in.\field{out}.\field{EPA}.\field{end}}} in.\field{out}.\field{EPA}.\field{power}  .
\end{align*}
The conditions for consumption and financial assets can be described
formally in similar ways.
\end{itemize}

\subsubsection{Smart-Meter Transactions}

Prosumers use smart-meter transactions to withdraw energy and
financial assets from their own smart meters, before they engage in
trading.
A transaction contains the following fields:
\begin{itemize}[noitemsep,topsep=-\parskip]
\item \field{EPA\_out}: list of EPA outputs (see above),
\item \field{ECA\_out}: list of ECA outputs (see above),
\item \field{FA\_out}: list of FA outputs (see above),
\item \field{id}: smart meter's identifier,
\item \field{sig}: smart meter's signature over the transaction.
\end{itemize}
\vspace{0.5\parskip} This transaction creates and transfers the assets
to the prosumer's addresses, which are specified in the output lists.

The smart meter signs the transaction only if the prosumer is allowed
to withdraw these assets.  More specifically, the amount of assets
withdrawn can never exceed certain limits that are set by the DSO.
For example, in the case of EPA, the following condition must be
satisfied for prosumer $i$:
\begin{equation*}
\forall t: \sum_{tr \,\in\, \field{STR}_i} \sum_{\substack{out \,\in\, tr.\field{EPA\_out}:\\out.\field{EPA}.\field{start} \leq t \leq out.\field{EPA}.\field{end}}} \hspace{-1em} out.\field{EPA}.\field{power} < \field{MAXEPA}_i ,
\end{equation*}
where $\field{STR}_i$ is the set of smart-meter transactions created
for prosumer $i$, and $\field{MAXEPA}_i$ is the withdrawal limit.  The
condition for consumption assets is similar, based on a withdrawal
limit $\field{MAXECA}_i$.  For financial assets, the smart meter can
take into account the amounts withdrawn and deposited, as well as the
outside bill payments to the DSO.

\Aron{To address malfunctioning or compromised smart meters, we could also impose a limit on withdrawals.}
A transaction is valid if the following two conditions hold.
\begin{itemize}[noitemsep,topsep=-\parskip]
\item The smart meter identified in the transaction has been authorized (and not been banned) by regulatory transactions. % that was previously recorded on the ledger.
\item The smart meter's signature is valid (for the smart meter's public key, see regulatory transactions).
\end{itemize}

\subsubsection{Regulatory Transactions}

The DSO uses regulatory transactions for two purposes: (1) to manage
the set of authorized smart meters and (2) to change the price policy.
First, whenever a new smart meter is installed, the DSO notifies the
microgrid by authorizing the device using a regulatory transaction.
Likewise, whenever a smart meter is deactivated (\emph{e.g.}, because
service is stopped or the device is believed to be faulty or
compromised), the DSO notifies the microgrid by banning the device.
Second, to influence microgrid load, the DSO can set a price policy,
which includes a price at which prosumers may buy energy from the DSO
and a price at which they may sell energy to the DSO.

A regulatory transaction contains the following fields:
\begin{itemize}[noitemsep,topsep=-\parskip]
\item \field{authorize}: list of smart meters to be authorized, each
  of which is defined by
\begin{compactitem}
\item \field{id}: identifier of the smart meter,
\item \field{pubkey}: public key of the smart meter,
\end{compactitem}
\item \field{ban}: list of identifiers of smart meters to be banned, 
\item \field{priceConsumption}: price at which DSO sells energy,
\item \field{priceProduction}: price at which DSO buys energy,
\item \field{time}: timestep after which authorizations, bans, and price changes should take effect,
\item \field{sig}: DSO's signature over the transaction.
\end{itemize}
\vspace{0.5\parskip}

A regulatory transaction of this type is valid if \field{timestep} is
not in the past and the DSO's signature is valid.  The active prices
for timestep $t$ are given by the last regulatory transaction recorded
on the ledger whose \field{time} is less than $t$.  Likewise,
regulatory transactions that are recorded on the ledger later override
the authorizations and bans of earlier transactions.

%!TEX root = paper.tex
\subsection{Services}

We now describe the various services that are provided in PETra.
Earlier, we discussed the distributed ledger, which permanently
stores valid transactions.  Below, we introduce the anonymous
communication service, the mixing service for transaction anonymity,
the anonymous bid storage, and smart-meter based billing.

\subsubsection{Communication Anonymity}
The anonymous communication layer is the infrastructure upon which all
other anonymity services in PETra are built.  Without this communication
layer, transactions and bids could be easily de-anonymized based on
their sources' network identifiers (\emph{e.g.}, IP or MAC addresses).

We can employ well-known and widely used techniques for anonymous
communication, such as \emph{onion routing}~\cite{reed1998anonymous}.
To build an onion network, the smart meters, prosumers, and other
devices can act as onion routers, and the list of onion routers in a
microgrid can be published on the ledger.  In practice, this service
can be built on the free and open-source Tor software with private
Directory Authorities.  In this case, anonymous communication
identifiers in bids and asks correspond to public-keys that identify
Tor hidden services.
% ritter.vg: "run your own tor network"
% https://ritter.vg/blog-run_your_own_tor_network.html

\subsubsection{Transaction Anonymity}
Communication anonymity is necessary, but not sufficient, for
anonymous trading. In particular, if prosumers used their own accounts
to transfer assets, their trades would not be anonymous.  Fortunately,
most distributed ledgers allow users to easily generate new addresses
at random, which are anonymous in the sense that no one can tell who
generated them.  If prosumers simply transferred assets to these
addresses, however, they could be easily de-anonymized by tracing the
assets back to the prosumers.

To prevent this de-anonymization, prosumers transfer assets to their
anonymous addresses through a \emph{mixing service}.  The mixing
service prevents tracing assets back to their original owners by
mixing together multiple incoming transfers and multiple outgoing
transfers. This service thus hides the connections between the
prosumers and the anonymous addresses.

A mixing service can be implemented using multiple approaches.  The
simplest one is to use a \emph{trusted third party}, called a
cryptocurrency tumbler, which can receive and send assets. Anonymity
in this case, however, depends on the trustworthiness and reliability
of the third party, who could easily de-anonymize the addresses.  A
more secure approach is to use decentralized protocols, such as
CoinShuffle~\cite{ruffing2014coinshuffle} or
Xim~\cite{bissias2014sybil}.  These protocols enable participants to
mix assets with each other, thereby eliminating the need for a trusted
third party.  Some newer cryptocurrencies, such as
Zerocoin~\cite{miers2013zerocoin}, provide built-in mixing services,
which are often based on cryptographic principles and proofs.

\subsubsection{Bidding Anonymity}
Prosumers must also be able to anonymously post energy bids and asks
on the bid storage service.  An anonymous bid (or ask) contains an ECA
(or EPA), a price, and an anonymous communication identifier
(\emph{e.g.}, Tor hidden service), which can be used to contact the
bidding (or asking) prosumer.  To enforce safety requirements, the bid
storage service must verify that the prosumer actually owns the asset
to be traded.  To this end, the prosumer first has to prove that it
controls the anonymous address where the asset is stored, which can be
performed in multiple ways.

In many distributed ledgers, an address represents a public key, and
controlling means knowing the corresponding private key.  In this
case, the prosumer can prove that it controls an address by signing a
challenge, which was freshly generated by the service, with the
private key of the address.  Alternatively, the prosumer may also
prove control by transferring zero amount of assets to a random
address that was freshly generated by the service.

\subsubsection{Smart-Meter Based Billing}
After a prosumer has finished trading, it deposits all of its EPA,
ECA, and FA to the smart meter by transferring them to an anonymous
address generated by the smart meter.  Later, during timeslot $t$, the
smart meter measures the amount of energy actually consumed (or
produced) by the prosumer using physical sensors.  The meter can then
compute the prosumer's bill for timeslot $t$, which will be paid to
the DSO, as follows. \Abhishek{The safety argument should be
  strengthened here}

The energy consumption balance $E_i^t$ of prosumer $i$ is
\begin{align*}
E_i^t = & \hphantom{+} \text{measured net energy consumption during timeslot } t \\ 
 & - \sum_{epa \,\in\, \{\text{EPA deposited by } i\}: ~ epa.\field{start} \,\leq\, t \,\leq\, epa.\field{end}} epa.\field{power} \\
 & + \sum_{epa \,\in\, \{\text{EPA withdrawn by } i\}: ~ epa.\field{start} \,\leq\, t \,\leq\, epa.\field{end}} epa.\field{power} .
\end{align*}
Notice that consumption assets are not used directly for billing, they are only used to enforce security and safety requirements.

The bill $B_i^t$ of prosumer $i$ for timeslot $t$, which will be paid
by the prosumer to the DSO, is
\begin{align*}
B_i^t = &  \text{FA withdrawn by $i$ during $t$} - \text{FA deposited by $i$ during $t$} \\
 & + \begin{cases}
- E_i^t \cdot \field{priceProduction} & \text{ if } E_i^t < 0 \\
 E_i^t \cdot \field{priceConsumption} & \text{ otherwise,} 
\end{cases}
\end{align*}
where {\field{priceProduction}} and {\field{priceConsumption}} are the
prices set by the latest regulatory transactions for timeslot $t$.

%!TEX root = paper.tex
\section{Discussion}
\label{sec:discussion}

This section presents a semi-formal analysis of PETra and shows that
it satisfies the security, safety, and privacy requirements formulated
earlier.

\subsection{Security}
Satisfaction of the security requirements follows from:
\begin{itemize}[noitemsep,topsep=-\parskip]
\item immutability of transactions in the distributed ledger,
\item validity conditions of the transactions, which include
  conditions on both signatures and asset balances,
\item and tamper-resistance of smart meters.
\end{itemize}
Together, these properties ensure that only the right entities may
create and sign a transaction, that transactions adhere to the rules
of the trading workflow, and that transactions cannot be tampered
with.\footnote{Due to lack of space, we leave a detailed discussion
  and proof for future work.}

\subsection{Safety}
We now demonstrate that faulty or malicious prosumers cannot trade
excessive amounts of energy if normal prosumers follow the rules of
the trading workflow.  First, we can show that the net amount of
energy sold by prosumer $i$ for each timestep is at most
$\field{MAXEPA}_i$.  Due to the rules of the trading workflow, the gross amount of energy sold is less than or equal to the amount of EPA
obtained by prosumer $i$.
A prosumer can obtain EPA either by withdrawing from its smart meter
or by purchasing from another prosumer.  From its smart meter,
prosumer $i$ can withdraw at most $\field{MAXEPA}_i$.  Although the
prosumer may also buy EPA from another prosumer, this constitutes
buying energy, which decreases the net amount of energy sold with the
same amount.  Hence, the net amount of energy sold by prosumer $i$
cannot exceed $\field{MAXEPA}_i$.  By extending the argument, we can
show that the net amount of energy sold by a group of prosumers~$G$
cannot exceed $\sum_{i \in G} \field{MAXEPA}_i$.  Similarly, %we can
%show that
 the net amount of energy bought by a group of prosumers $G$
cannot exceed $\sum_{i \in G} \field{MAXECA}_i$.

%Using a similar argument, we can also show that the total amount of energy bids (or asks) posted at
%the same time by prosumer $i$ for each timestep is at most
%$\field{MAXEPA}_i + \field{MAXECA}_i$.  This limit is higher than for
%net energy sold or bought, since prosumer $i$ may purchase
%$\field{MAXECA}_i$ amount of EPA (or $\field{MAXEPA}_i$ amount of ECA)
%from other prosumers, and then post an energy ask (or bid) in the
%amount of $\field{MAXEPA}_i + \field{MAXECA}_i$.

\subsection{Privacy}
Due to our use of communication anonymity and mixing services, members
of a microgrid can observe only the amount of assets withdrawn by a
prosumer from its smart meter.  Since all trading transactions are
anonymous, they do not reveal the actual amount of assets traded by
the prosumer.  If a prosumer has not traded away all of its assets,
then it can also anonymously deposit the remainder to a random address
that was freshly generated by its smart meter. Even if a prosumer does
not wish to trade, it should always withdraw, mix, and deposit the
same amount of assets.  Otherwise, the lack (or varying amount) of
withdrawal would leak information.

As for the DSO, it receives the same information from the smart meter
as in a non-transactive smart grid (\emph{i.e.}, amount of energy
produced and consumed).  Since trading is anonymous, the DSO learns
only the financial balance of the prosumer, which is necessary for
billing.  However, we can provide an even higher-level of privacy.  In
particular, since price policies are recorded on the ledger (which the
smart meters may read), each prosumer's smart meter may calculate and
send the prosumer's monthly bill to the DSO, without revealing the
prosumer's energy consumption or production.  Meanwhile, the DSO can
still obtain detailed load information (including predictions) for the
microgrid from the bid storage and the trades recorded on the ledger.

%!TEX root = paper.tex
\section{Related Work}
\label{sec:related}

%In this section we describe the related research. It has been divided into two subsections to focus on blockchains, which we are using to implement the distributed ledger service described earlier, and other works on smart grid privacy concerns. 

%\subsection{Smart Grid and Meter Privacy}
% http://ieeexplore.ieee.org/abstract/document/5054916/
%McDaniel and McLaughlin discuss privacy challenges in smart grids~\cite{mcdaniel2009security}.
% https://arxiv.org/pdf/1108.2234.pdf
% https://pdfs.semanticscholar.org/cdf8/a5b6256823bca38a1d2347ab36f8e4a2ca94.pdf
% http://www.comm.toronto.edu/~akhisti/sm.pdf
% https://www.researchgate.net/profile/Georgios_Kalogridis/publication/224189766_Smart_Grid_Privacy_via_Anonymization_of_Smart_Metering_Data/links/541169510cf2b4da1bec4193.pdf
% https://arxiv.org/pdf/1305.0735.pdf
New privacy concerns arise with the continuing adoption of smart
grids. In addition to old and new security threats (such as energy
theft and smart-meter malware), McDaniel and McLaughlin discuss the
privacy concerns of energy usage profiling that smart grids could
potentially enable~\cite{mcdaniel2009security}. Several approaches
have been investigated as potential means to provide privacy
protections for smart grid users.

Some approaches look to the use of protocols and/or frameworks to help
protect privacy. Rajagopalan et al.\ use tools from information theory
to present a framework that abstracts both the privacy and the utility
requirements of smart-meter
data~\cite{rajagopalan2011smart,sankar2013smart}. Their framework
leads to a novel tractable privacy-utility tradeoff problem with
minimal assumptions. Efthymiou and Kalogridis describe a method for
securely anonymizing frequent electrical metering data sent by a smart
meter~\cite{efthymiou2010smart}. Their approach is based on the
observation that frequent metering data may be required by an energy
distribution network for operational reasons, but it may not
necessarily need to be attributable to a specific smart meter. The
authors describe a method that provides a third-party escrow mechanism
for authenticated anonymous meter readings, which are hard to
associate with a particular smart meter.

Other approaches, such as additional hardware components, have also been explored 
for potential privacy gains. 
Varodayan and Khisti study using a
rechargeable battery for partially protecting the privacy of
information contained in a household's electrical load
profile~\cite{varodayan2011smart}. They show that stochastic battery
policies may leak 26\% less information than a best-effort policy,
which holds the output load constant whenever possible. Tan et
al.\ study privacy in a smart metering system from an information
theoretic perspective in the presence of energy harvesting and storage
units~\cite{tan2013increasing}. They show that energy harvesting
provides increased privacy by diversifying the energy source, while a
storage device can be used to increase both energy efficiency and
privacy.
% They show that there exists a trade-off between the information
% leakage rate and the wasted energy rate, and study the impact of the
% energy harvesting rate and the size of the storage device on this trade-off.

PETra extends this work by (1) leveraging a decentralized IoT
system for transactive energy and (2) addressing the novel privacy
threat posed by trading. In particular, while earlier work protected
the prosumers' privacy from the DSO, PETra also protects it from other
prosumers, as well as outside attackers.

A key element of PETra is its ability to distribute information among
peers via blockchains.  As blockchain technology develops and matures,
new frameworks, services, and protocols are being developed to
leverage the distributed ledgers provided by blockchains. For example,
Hyperledger Fabric is a platform for distributed ledger solutions,
which was designed to support pluggable implementations of different
components~\cite{hyperledger2017fabric}.
%Bitcoin Lightning Network is
%decentralized system, in which transactions are sent over a network of
%micropayment channels, whose transfer of value occurs
%off-blockchain~\cite{poon2016bitcoin}. 
Since this paper focuses on the theoretical foundations of PETra, any
of these  ledgers provide the required capabilities.

%\Abhishek{Aron this section should end with a few sentence about how our approach fits in. Perhaps just a few sentences reworded from introduction will be sufficient}

\iffalse
\subsection{Blockchains as Distributed Ledgers}

As Blockchain technology continues to develop and mature, new
frameworks, services, and protocols are being developed to leverage
blockchain's distributed ledger. Microsoft offers Blockchain as a
Service (BaaS) on Azure. Additionally, Microsoft has Project Bletchley
as its architectural approach to building an Enterprise Consortium
Blockchain Ecosystem, introducing two new concepts: blockchain
middleware and a secure means for calling code or data outside a
SmartContract or blockchain called
cryptlets~\cite{gray2016introducing}. Hyperledger Fabric is a platform
for distributed ledger solutions, which was designed to support
pluggable implementations of different
components~\cite{hyperledger2017fabric}. Bitcoin Lightning Network is
decentralized system, in which transactions are sent over a network of
micropayment channels whose transfer of value occurs
off-blockchain~\cite{poon2016bitcoin}. Interledger is a protocol for
payments across payment systems, which enables anyone with accounts on
two ledgers to create a connection between
them~\cite{thomas_protocol}. Xu et al.\ discuss the use of two pools
of proxy agents, an agreement pool and a payment pool, to assist in
protection of privacy when using blockchain technologies for
transactions on tangible goods~\cite{Xu2017}.  \fi

%\url{https://geli.net/residential/}
%\url{https://www.greentechmedia.com/articles/read/geli-raises-7m-to-take-energy-storage-software-to-the-next-level}

%\url{http://ethembedded.com/}

%!TEX root = paper.tex
\section{Concluding Remarks}
\label{sec:concl}

\Doug{Abhishek, as always, it would be best if we could revise this summary of the paper's contents to be a summary of lessons learned based upon our experiences to date!}
As the complexity of power systems increases due to the evolution of
power grids, decentralized transactive-energy IoT systems are emerging
to tackle this complexity. Ironically, these decentralized systems
also give rise to new privacy challenges, such as the potential leakage of energy usage patterns, including the possibility of inferring the future state of a prosumer.
These challenges are exacerbated by the stringent safety and
security requirements of power systems.

This paper describes 
\emph{Privacy-preserving Energy Transactions} (PETra), our
innovative solution for anonymous energy trading within a transactive
microgrid.  PETra builds on distributed ledgers, such as blockchains,
and proven techniques for anonymity, such as mixing services and onion
routing.  We described the workflow of anonymous energy trading and
explained the novel transactions and services used in PETra.  Finally,
we discussed how PETra satisfies security, safety, and privacy
requirements.  In future work, we will provide rigorous proofs of
satisfying these requirements.
% and we will present an implementation
%of PETra with performance results.

\section*{Acknowledgment}
This work was funded in part by a grant by Siemens Corporation, CT. 

% BALANCE COLUMNS
%\balance{}
%\setlength{\bibsep}{0pt plus 0ex}
%\let\oldthebibliography\thebibliography
%\let\endoldthebibliography\endthebibliography
%\renewenvironment{thebibliography}[1]{
%  \begin{oldthebibliography}{#1}
%    \setlength{\itemsep}{0.1em}
%    \setlength{\parskip}{0em}
%}
%{
%  \end{oldthebibliography}
%}

% REFERENCES FORMAT
% References must be the same font size as other body text.
\bibliographystyle{SIGCHI-Reference-Format}

\bibliography{references}

\end{document}